\documentclass[aps,pre,twocolumn,superscriptaddress,floatfix]{revtex4}
\usepackage{amssymb,amsfonts,amsmath}

\usepackage{color}
\usepackage{graphicx}

\newcommand{\ie}{{\it i.e.\ }}
\renewcommand{\emph}[1]{{\it #1}}

\begin{document}

\title{Landscape encodings enhance optimization}

\author{Konstantin Klemm}
\affiliation{Bioinformatics Group, Department of Computer Science, and
Interdisciplinary Center for Bioinformatics, University of Leipzig,
D-04107 Leipzig, Germany}

\author{Anita Mehta}
\affiliation{S N Bose National Centre for Basic Sciences, Calcutta 700 098,
 India}

\author{Peter F. Stadler}
\affiliation{Bioinformatics Group, Department of Computer Science, and
Interdisciplinary Center for Bioinformatics, University of Leipzig,
D-04107 Leipzig, Germany}
\affiliation{Max Planck Institute for Mathematics in the Science,
Inselstra{\ss}e 22, D-04103 Leipzig, Germany}
\affiliation{Fraunhofer Institut f{\"u}r Zelltherapie und Immunologie
-- IZI Perlickstra{\ss}e 1, D-04103 Leipzig, Germany}
\affiliation{Department of Theoretical Chemistry
University of Vienna,
W{\"a}hringerstra{\ss}e 17, A-1090 Wien, Austria}
\affiliation{Santa Fe Institute, 1399 Hyde Park Rd., Santa Fe NM 87501, USA
}

\date{today}

% Please keep the abstract between 250 and 300 words
\begin{abstract}
Hard combinatorial optimization problems deal with the
search for the minimum cost solutions (ground states)
of discrete systems under strong constraints. A
transformation of state variables
may enhance computational tractability. It has been
argued that these state encodings are to be chosen
invertible to retain the original size of the state
space.  Here we show how redundant non-invertible
encodings enhance optimization by enriching the density
of low-energy states. In addition, smooth landscapes may
be established on encoded state spaces to guide local
search dynamics towards the ground state.
\end{abstract}

\maketitle

\section{Introduction}

Complex systems in our world are often \emph{computationally} complex as
well.  In particular, the class of NP-complete problems \cite{Garey:1979a},
for which no fast solvers are known, encompasses not only a wide
  variety of well-known combinatorial optimization problems from the
  Travelling Salesman Problem to graph coloring, but also includes a rich
  diversity of applications in the natural sciences ranging from genetic
networks \cite{Berg:04} through protein folding \cite{Fraenkel:93} to spin
glasses \cite{Mezard:02,Bauer:05, Schulman:07, Ricci-Tersenghi:2010}.  In
such cases, heuristic optimization -- where the goal is to find the best
solution that is reachable within an allocated time -- is widely accepted
as being a more fruitful avenue of research than attempting to find an
exact, globally optimal, solution. This view is motivated at least in part
by the realization that in physical and biological systems, there are
severe constraints on the type of algorithms that can be naturally
implemented as dynamical processes.  Typically, thus, we have to deal with
local search algorithms. Simulated annealing \cite{Kirkpatrick:1983},
genetic and evolutionary algorithms \cite{Holland:1992}, as well as genetic
programming \cite{Koza:1992} are the most prominent representatives of this
type. Their common principle is the generation of variation by thermal or
mutational noise, and the subsequent selection of variants that are
advantageous in terms of energy or fitness \cite{Reidys:02a}.

The performance of such local search heuristics naturally depends on the
structure of the search space, which, in turn, depends on two ingredients:
(1) the encoding of the configurations and (2) a move set. Many
combinatorial optimization problems as well as their counterparts in
statistical physics, such as spin glass models, admit a natural encoding
that is (essentially) free of redundancy. In the evolutionary computation
literature this ``direct encoding'' is often referred to as the ``phenotype
space'', $X$. The complexity of optimizing a cost function
$f$ over $X$ is determined already at this level. For simplicity, we call
$f$ energy and refer to its global minima as ground states. In
evolutionary computation, one often uses an additional encoding $Y$, called
the ``genotype space'' on which search operators, such as mutation and
cross-over, are defined more conveniently \cite{Rothlauf:03,Rothlauf:06}.
The genotype-phenotype relation is determined by a map $\alpha:Y\to
X\cup\{\varnothing\}$, where $\varnothing$ represents phenotypic
configurations that do not occur in the original problem, \ie non-feasible
solutions. For example, the
tours of a Traveling Salesman Problem (TSP) \cite{TSP:06} are directly
encoded as permutations describing the order of the cities along the
tour. A frequently used encoding as binary strings represents every
connection between cities as a bit that can be present or absent in a tour;
of course, most binary strings do not refer to valid tours in this picture.

The move set (or more generally the search operators \cite{Flamm:07a})
define a notion of locality on $X$. Here we are interested only in
mutation-based search, where for each $x\in X$ there is a set of neighbors
$N(x)$ that is reachable in a single step. Such neighboring configurations
are said to be \emph{neutral} if they have the same fitness. Detailed
investigations of fitness landscapes arising from molecular biology have
led to the conclusion that high degrees of neutrality \emph{can} facilitate
optimization \cite{Schuster:94a,Reidys:02a}.  More precisely, when
populations are trapped in a metastable phenotypic state, they are most
likely to escape by crossing an entropy barrier, along long neutral paths
that traverse large portions of genotype space \cite{vanNimwegen:00}.

In contrast, some authors advocate to use ``synonymous encodings'' for
  the design of evolutionary algorithms, where genotypes mapping to the
  same phenotype $x\in X$ are very similar, i.e., $\alpha^{-1}(x)$ forms a
  local ``cluster'' in $Y$, see e.g.\
  \cite{Rothlauf:06,Choi:08,Rothlauf:11}.  This picture is incompatible
  with the advantages of extensive neutral paths observed in biologically
  inspired landscape models \cite{Schuster:94a,Fernandez:07} and in genetic
  programming \cite{Yu:02,Banzhaf:06}.  An empirical study
\cite{Knowles:02}, furthermore, shows that the introduction of arbitrary
redundancy (by means of random Boolean network mapping) does not increase
the performance of mutation-based search. This observation can be
understood in terms of a random graph model of neutral networks, in which
only very high levels of randomized redundancy result in the emergence of
neutral paths \cite{Reidys:97a}.

An important feature that appears to have been overlooked in most recent literature 
is that the redundancy of $Y$ with respect to $X$ need not be homogeneous
\cite{Rothlauf:03}. Inhomogeneous redundancy implies that the size of the
preimage $|\alpha^{-1}(x)|$ may depend on $x \in X$. If $|\alpha^{-1}(x)|$
is anti-correlated with the energy $f(x)$, then the encoding $Y$ enables
the preferential sampling of low-energy states in $X$. Thus even a random
selection of a state yields lower energy when performed in $Y$ than in
$X$. Here we demonstrate this \emph{enrichment} of low energy states for
three established combinatorial optimization problems and suitably chosen
encodings. The necessary formal aspects of energy landscapes and their
encodings are outlined in the Methods section. We formalize and measure
enrichment in terms of densities of states on $X$ and $Y$, see
  \emph{Methods} for a formal treatment. We illustrate the effects of
encoding by comparing performance of optimization heuristics on the direct
and encoded landscapes.

\section{Results and Discussion}

\subsection{Number Partitioning}

The first optimization problem we consider is the number partitioning
problem (NPP) \cite{Garey:1979a}: this asks if one can divide $n$ positive
numbers $a_1,a_2,\dots,a_n$ into two subsets such that the sum of elements
in the first subset is the same as the sum over elements in the other
subset. The energy is defined as the deviation from equal sums in the two
subsets, i.e., 
\begin{equation}
f(x) = \left| \sum_{i=1}^n x_i a_i \right| 
\end{equation}
where the two choices
$x_i \in \{-1,+1\}$ correspond to assignment to the first or to the second
subset, respectively. The flipping of one of the spin variables $x_i$ is
used as a move set, so that the NPP landscape is built on a
hypercube. The NPP shows a phase transition between an easy and
a hard phase. We consider here only instances that are hard in practice,
i.e., where the coefficients $a_i$ have a sufficiently large number of
digits \cite{Mertens:1998}.

The so-called \emph{prepartitioning} encoding \cite{Ruml:1996} of the NPP
is based on the differencing heuristic by Karmakar and Karp
\cite{Karmakar:82}.  Departing from an NPP instance $(a_1,\dots,a_n)$, the
heuristic removes the largest number, say $a_i$, and the second largest
$a_j$ and replaces them by their difference $a_i-a_j$. This reduces the
problem size from $n$ to $n-1$. After iterating this differencing step
$n-1$ times, the single remaining number is an upper bound for --- and in
many cases a good approximation to --- the global minimum energy. The
minimizing configuration itself is obtained by keeping track of the items
chosen for differencing. Replacing $a_i$ and $a_j$ by their difference
amounts to putting $a_i$ and $a_j$ into different subsets, i.e.\ $x_i \neq
x_j$.

The prepartitioning encoding is obtained by modifying the initial condition
of the heuristic. Each number $a_i$ is assigned a class $y_i \in
\{1,\dots,n\}$. A new NPP instance $a^\prime_1, \dots,a^\prime_{n}$ is
generated by adding up all numbers $a_i$ in the same class $y_i$ into a
single number $a^\prime_{y_i}$.  After removing zeros from $a^\prime$, the
differencing heuristic is applied to $a^\prime$. In short: $y_i = y_j$
imposes the constraint $x_i = x_j$. Running the heuristic under this
constraint, the resulting configuration $x=\alpha(y)$ is unique up to
flipping all spins in $x$.  The so defined mapping $\alpha: Y \rightarrow
X$ is surjective because for each $x \in X$, $\alpha(y) =x$ for $y_i=1$ if
$x_i=1$ and $y_i=2$ otherwise. Two encodings $y,z \in Y$ are neighbors if
they differ at exactly one index $i \in \{1,\dots,n\}$. This encoding is
the one whose performance we will compare with the direct encoding later.

\subsection{Traveling Salesman}

Our next optimization problem, the Traveling Salesman Problem, (TSP) is
another classical NP-hard optimization problem \cite{Garey:1979a}. Given a
set of $n$ vertices (cities, locations) $\{1,\dots,n\}$ and a symmetric
matrix of distances or travel costs $d_{ij}$, the task is to find a
permutation (tour) $\pi$ that minimizes the total travel cost
\begin{equation}
f(\pi) = \sum_{i=1}^n d_{\pi(i),\pi(i+1)}
\end{equation}
where indices are interpreted modulo $n$. Here, the states of the landscape
are the permutations of $\{1,\dots,n\}$, $X=S_n$. Two permutations $\pi$
and $\sigma$ are adjacent, $\{ \pi,\sigma \} \in L$, if they differ by one
\emph{reversal}.  This means that there are indices $i$ and $j$ with $i<j$
such that $\sigma_k = \pi_{i+j-k}$ for $i \le k \le j$ and 
$\sigma_k = \pi_k$ otherwise.

Similar to the NPP case, an encoding configuration $y \in
Y:=\{1,2,\dots,n\}^n$ acts as a constraint.  A tour $\pi \in X$ fulfills
$y$ if for all cities $i$ and $j$, $y_i \le y_j$ implies $\pi^{-1}(i) \le
\pi^{-1}(j)$.  Thus $y_i$ is the relative position of city $i$ in the tour
since it must come after all cities $j$ with $y_j<y_i$. All cities with the
same $y$-value appear in a single section along the tour. If there are no
two cities with the same $y$-value then $y$ itself is a permutation and
there is a unique $\pi \in X$ obeying $y$, namely $\pi = y^{-1}$.

Among the tours compatible with the constraint, a selection is made with
the greedy algorithm. It constructs a tour by iteratively fixing
adjacencies of cities. Starting from an empty set of adjacencies, we
attempt to include an adjacency $\{i,j\}$ at each step. If the resulting
set of adjacencies is still a subset of a valid tour obeying the
constraint, the addition is accepted, otherwise $\{i,j\}$ is discarded. The
step is iterated, proposing each $\{i,j\}$ exactly once in the order of
decreasing $d_{i,j}$. This procedure establishes a mapping (encoding)
$\alpha:Y \rightarrow X$. Since each tour $\pi$ can be reached by taking
$y=\pi^{-1}$, $\alpha$ is complete. In the encoded landscape, two states
$y,z \in Y$ are adjacent if they differ at exactly one position (city) $i$.

\subsection{Maximum Cut}

The last example we consider is a Spin Glass problem.
Consider the set of configurations $X=\{-1,+1\}^n$ with the energy function
\begin{equation}
f(x) = - \sum_{i,j} J_{ij} x_i x_j
\end{equation}
for a spin configuration $x \in X$. Proceeding differently from the usual
Gaussian or $\pm J$ spin glass models \cite{Sherrington:1975,Binder:1986},
we allow the coupling to be either antiferromagnetic or zero, $J_{ij} \in
\{-1,0\}$. This is sufficient to create frustration and obtain hard
optimization problems. Taking the negative coupling matrix $-J$ as the
adjacency matrix of a graph $G$, the spin glass problem is equivalent to
the max-cut problem on $G$, which asks to divide the node set of $G$ into
two subsets such that a maximum number of edges runs between the two
subsets \cite{Garey:1979a}.

The idea for an encoding works on the level of the graph $G$, which we
assume to be connected. The set $Y$ of the encoding consists of all
spanning trees of $G$. In the mapped configuration $x=\alpha(y)$, $x_i$ and
$x_j$ have different spin values whenever $ij$ is an edge of the spanning
tree $y$.  Since a spanning tree is a connected bipartite graph, this
uniquely (up to $+1/-1$ symmetry) defines the spin configuration $x$. The
encoding $\alpha$ is not complete in general. Homogeneous spin
configurations, for instance, are not generated by any spanning tree. Each
ground state configuration $x_\text{ground}$, however, is certain to be
represented by a spanning tree due to the following argument.
Suppose there is a minimum energy configuration $x_\text{ground}$ that is
not generated by any spanning tree. Then the subgraph of $G$ formed by all edges
connecting unequal spins in $x_\text{ground}$ is disconnected. We choose
one of the connected components, calling its node set $C$. By
flipping all spins in $C$, we keep all edges present for $x_\text{ground}$.
Since $G$ is connected, we obtain at least one additional edge from a node in
$C$ to a node outside $C$. Thus we have constructed a configuration with
strictly lower energy than $x_\text{ground}$, a contradiction.
Two spanning trees $y,z \in Y$ are adjacent, if $z$ can be obtained from $y$ by
addition of an edge $e$ and removal of a different edge $f$.

%%%%%%%%%%%%%%%%%%%%%%%%%%%%%%%%%%%%%%%%%%%%%%%%%%%%%%%%%%%%%%%%%%%%%%%%%%%%%
\begin{figure*}
\begin{center}
\includegraphics[width=\textwidth]{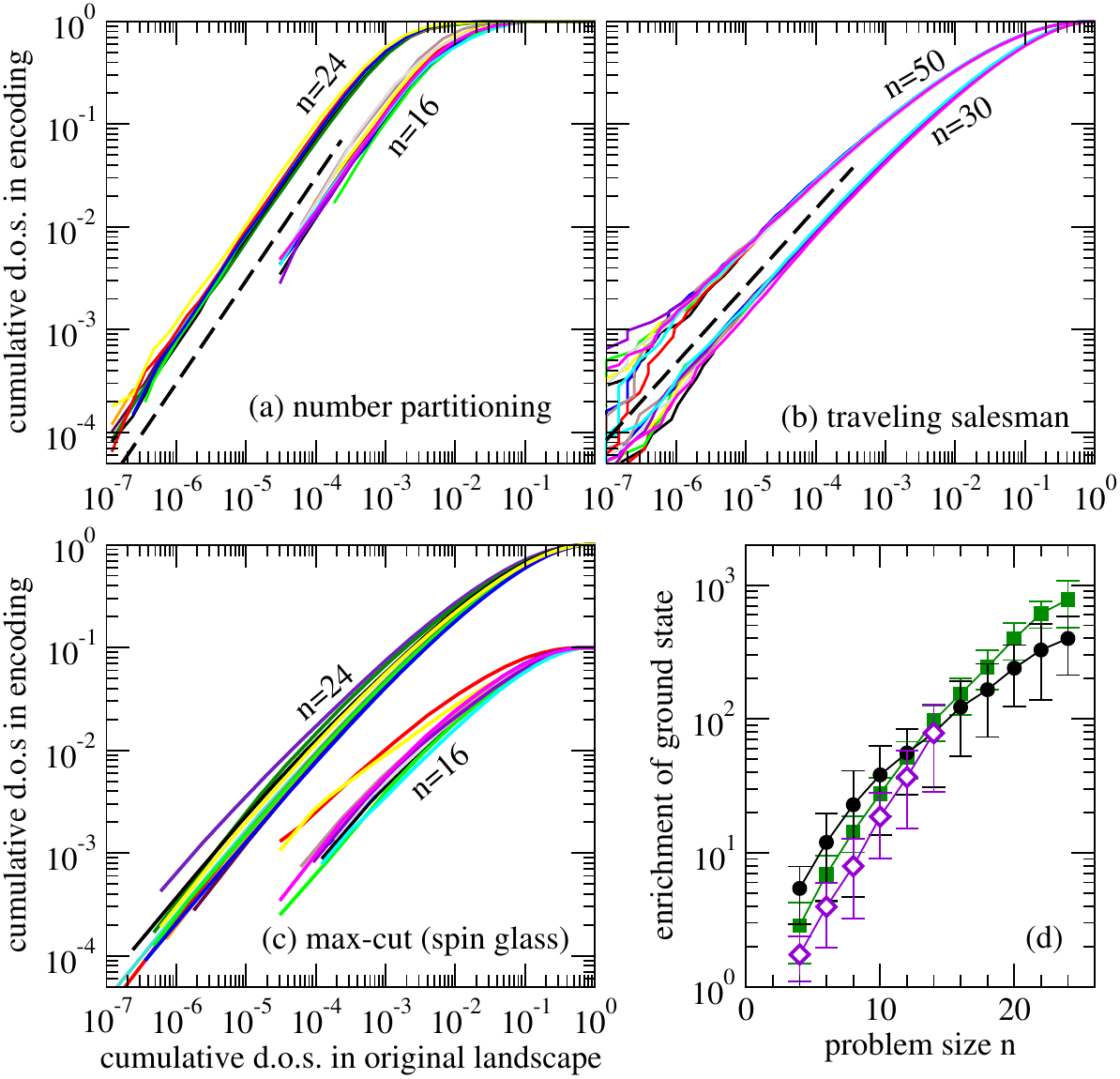}
\end{center}
\vspace*{3mm}
\caption{\label{fig:enrichment} {\bf Enrichment of the density of low energy states
  for landscape encodings}. In panels (a,b,c), a point $(h,r(h))$ on a curve
  indicates a fraction $h$ of all states have an energy not larger than a
  certain threshold $\eta$ in the original landscape whereas this fraction is
  $r(h)$ using the same energy threshold in the encoding. Panel (d) shows the
  average enrichment of the ground state as a function of problem size for
  traveling salesman ($\diamondsuit$), number partitioning ($\Box$), and
  max-cut ($\circ$). Error bars give the standard deviation
  over 100 independent realizations. In panels (a-c), the solid curves are for
  10 random instances of each landscape and system size. The dashed lines
  follow $r(h) \propto h$  in panel (a) and $r(h) \propto h^{3/4}$ in panel
  (b).}
\end{figure*}

\subsection{Enrichment}

We now study enrichment as well as landscape structure on these three
rather different problems. To this end, we consider the cumulative density
of states
\begin{equation}
Q_f(\eta) = | \{ x \in X : f(x)\le \eta\}| \, / \, |X|
\end{equation}
in the original landscape and $Q_{f\circ \alpha}$ defined analogously in
the encoded landscape. In order to quantify the enrichment of good
  solutions, we compare the fraction $h$ of all states with an energy not
  larger than a certain threshold $\eta$ in the original landscape with the
  fraction $r(h)$ using the same threshold in the encoding.  The encoding
thus enriches low energy states if $r(h) \gg h$ for small $h$. Figure
\ref{fig:enrichment} shows that this is the case for the three landscapes
and encodings considered here. We find in fact that the density of states
$r(h)/h$ is enriched by several orders of magnitude in the encoded
landscape, for all the cases considered.

Reassuringly, this trend of enrichment persists all the way to the ground
state: that is, the encodings contain many more copies of the ground state
than the original landscape. It appears in fact that the enrichment
of ground states increases exponentially with system size. We can
thus conclude that with the choice of an appropriately encoded landscape, it
is easier \emph{both} to find lower energy states from higher energy ones,
and thus have more routes to travel to the ground state, as well as to
\emph{reach} the ground state itself from a low-energy neighbor, as a
result of enrichment.

%%%%%%%%%%%%%%%%%%%%%%%%%%%%%%%%%%%%%%%%%%%%%%%%%%%%%%%%%%%%%%%%%%%%%%%%

\subsection{Neighborhoods and neutrality}

\begin{figure*}
\begin{center}
\includegraphics[width=\textwidth]{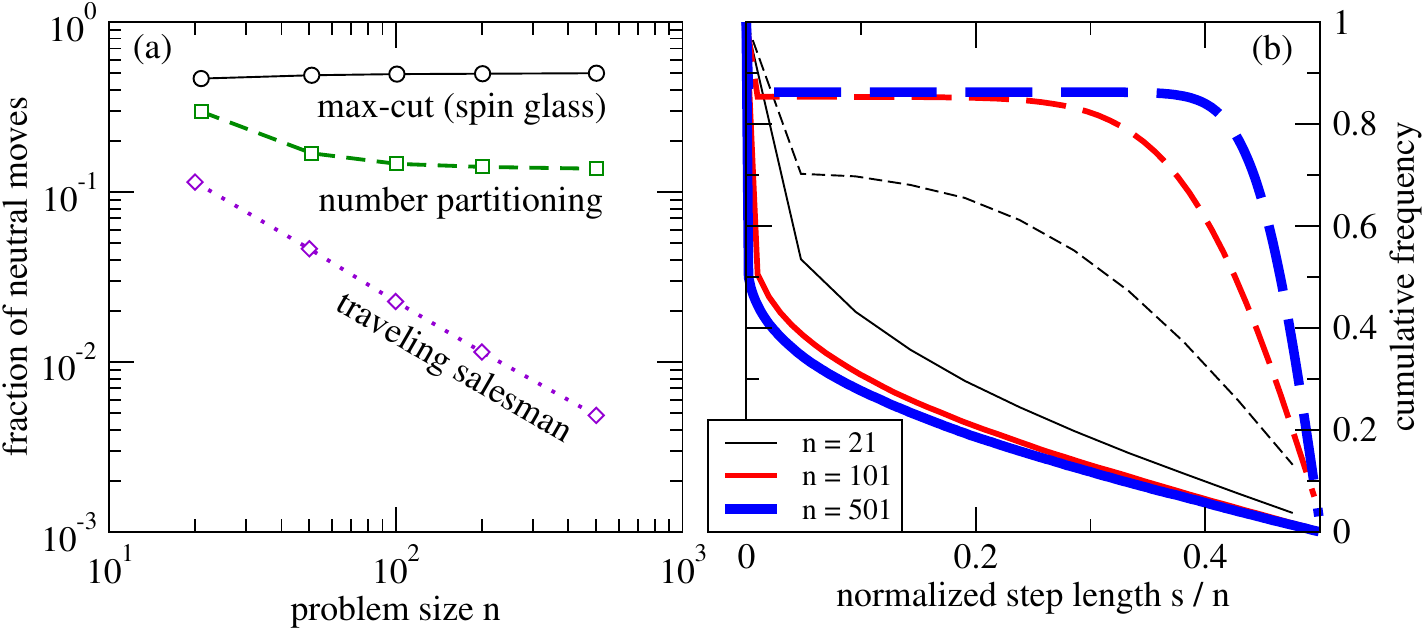}
\end{center}
\vspace*{3mm}
\caption{\label{fig:neigh} {\bf Neutrality and encoded step length}.
(a) The fraction of neutral neighbors as a function of problem size.
(b) The cumulative distribution of the distance moved in the original
landscape by a single step in the encoding. Solid curves are for the
max-cut, dashed curves for the number partitioning problem, with curve
thickness distinguishing values of problem size $n$. For both plots
(a) and (b), data have been obtained by uniform sampling of 
$10^4$ neighboring state pairs on
$10^2$ independently generated instances of each type of landscape.
}
\end{figure*}

We continue the analysis of the encodings with attention to geometry and
distances. A \emph{neutral} mutation is a small change in the genotype that
leaves the phenotype unaltered. In the present setting, a neutral move in
the encoding is an edge $\{y,z\} \in M$ such that $\alpha(y) =
\alpha(z)$. In general, the set of neutral moves is a subclass of all moves
leaving the energy unchanged. An edge $\{x,y\}$ with $f(\alpha(x))=f(\alpha(y))$
but $\alpha(x)\neq \alpha(y)$ is not a neutral move in the present context.
In the following, we examine the fraction of neutral moves for
the encoded landscapes mentioned above.

Figure~\ref{fig:neigh}(a) shows that the fraction of neutral moves approaches
a constant value when increasing the problem size of NPP and max-cut.  
The fraction of neutral moves in the traveling
salesman problem, on the other hand, decreases as $1/n$ with problem size
$n$. The average number of neighbors encoding the same solution grows
linearly with $n$, since the total number of neighbors is $n(n-1)$ for
each $y \in Y$ in the TSP encoding.

If a move in the encoding is non-neutral, how far does it take us on the
original landscape? We define the step length of a move $\{y,z\} \in Y$ as 
the distance between the images of $y$ and $z$ on the original landscape,
\begin{equation}
s(\{y,z\}) = d_X(\alpha(y),\alpha(z))
\end{equation}
using the standard metric $d_X$ on the graph $(X,L)$. Obviously, $\{y,z\}$
is neutral if and only if $s(\{y,z\})=0$. Figure~\ref{fig:neigh}(b)
compares the cumulative distributions of step length for number
partitioning and max-cut. It is intractable to get the statistics of $s$
for the TSP problem for larger problem sizes since sorting by reversals,
i.e., measuring distances w.r.t.\ to the natural move set, is a known
NP-hard problem \cite{Caprara:97}.

For the encoding of number partitioning, step lengths are concentrated
around $n/2$. Making a non-neutral move in this encoding is therefore akin
to choosing a successor state at random. For the max-cut problem, the
result is qualitatively different. Step lengths are broadly distributed
with most moves spanning a short distance on the original landscape. Based
on this it is tempting to conclude that optimization proceeds in 'smaller
steps' on the max-cut landscape, than in the NPP problem.

%%%%%%%%%%%%%%%%%%%%%%%%%%%%%%%%%%%%%%%%%%%%%%%%%%%%%%%%%%%%%%%%%%%%%%%%%%%%

\subsection{Evolutionary dynamics}

\begin{figure*}
\begin{center}
\includegraphics[width=\textwidth]{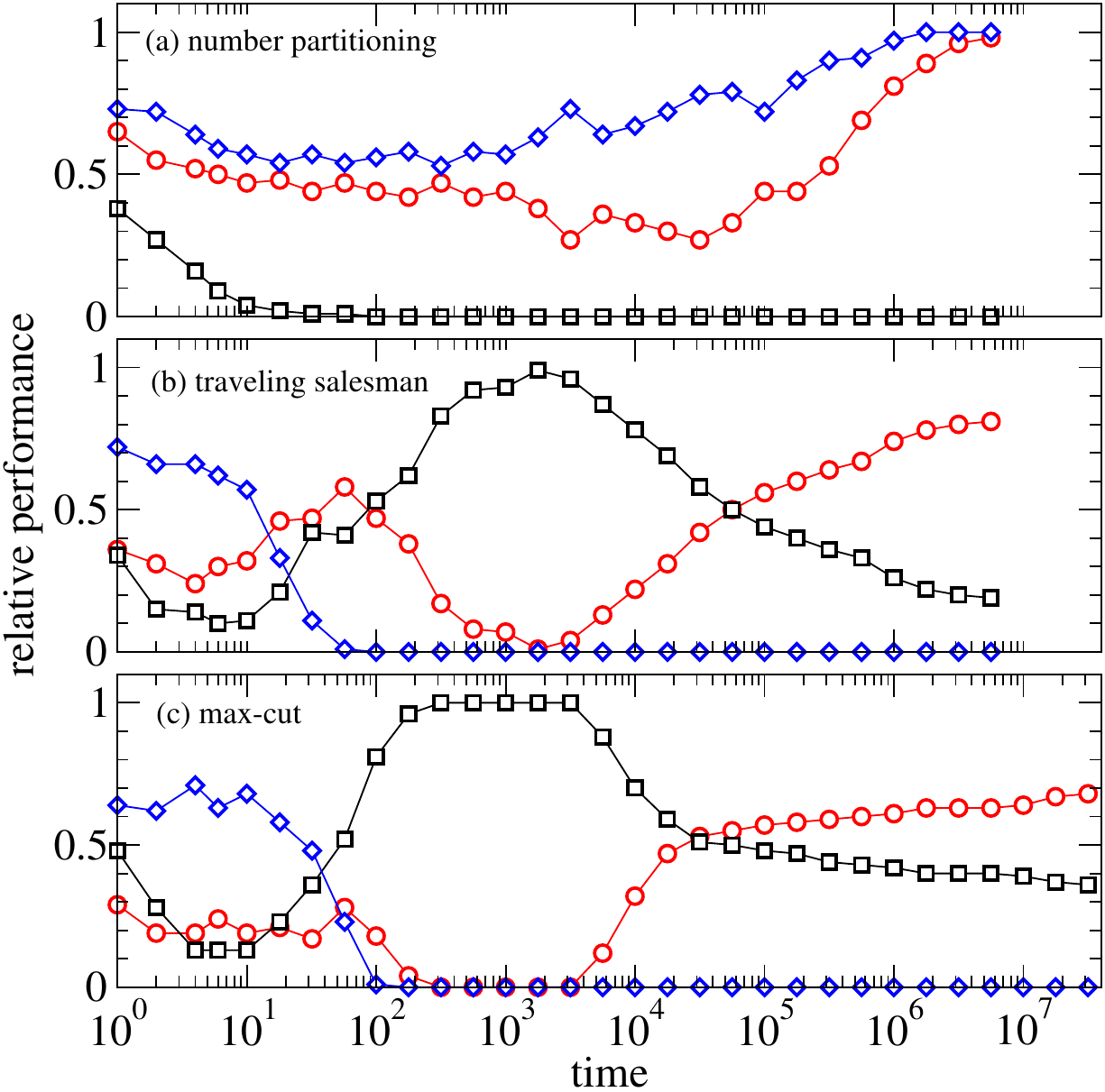}
\end{center}
\vspace*{3mm}
\caption{\label{fig:dynamics} {\bf Performance comparison between three types of
  stochastic dynamics}: adaptive walks (AW) on the original ($\square$) and
  encoded ($\circ$) landscapes and randomly generate and test (RGT) on the
  encoded landscape ($\diamond$). The plotted performance value is the
  fraction of instances for which the considered evolutionary dynamics is
  ``leading'' at time $t$, i.e.\ has an energy not larger than the other
  two types of dynamics. 
  For each landscape, 100
  random instances are used with sizes $n=30$ in panels (a) and (b), 
  $n=200$ in panel (c). On each of the instances,
  each type of evolutionary dynamics is run once with randomly drawn
  initial condition $y(0) \in Y$ for RGT and AW in the encoded
  landscape. The AW on the original landscape is initialized with the
  mapped state $x(0)=\alpha(y(0))$. Thus all three dynamics are started at
  the same energy.  }
\end{figure*}

One might ask if the encoded landscape also facilitates the search
dynamics, by virtue of its modified structure, and offers another avenue
for optimization.  For this purpose, we consider an optimization dynamics
as a zero-temperature Markov chain $x(0),x(1),x(2),\dots$. At each time
step $t$, a proposal $x^\prime$ is drawn at random. If $f(x^\prime)\le
x(t)$, we set $x(t+1)=x^\prime$, otherwise $x(t+1)=x(t)$. This is an
Adaptive Walk (AW) when the proposal $x^\prime$ is drawn from the
neighborhood of $x(t)$. In Randomly Generate and Test (RGT), proposals are
drawn from the whole set of configurations independently of the
neighborhood structure. Thus a performance comparison between AW and RGT
elucidates if the move set is suitably chosen for optimization. Because of
the enrichment of low energy states by the encodings, it is clear that RGT
performs strictly better on the encoding than on the original landscape.

Adaptive walks also perform strictly better on the encoding than on the
original landscape, at least in the long-time limit, cf.\
Figure~\ref{fig:dynamics}.  Beyond this general benefit of the encodings,
the dynamics shows marked differences across the three optimization
problems. In the NPP problem, RGT outperforms AW on the encoded landscape,
so that enrichment alone is responsible for the increase in optimization
with respect to the original landscape. In the encodings of the other two
problems, AW performs better than RGT so that we can conclude that the
improved structure of the encoded landscape is also an important reason for
the observed increase in performance, in addition to simple enrichment. The
dynamics on the max-cut landscapes (panel c) has the same qualitative
behavior as that on the TSP (panel a). Although there is a transient for
intermediate times where adaptive walks on the original landscape seem to
be winning, the asymptotic behavior is clear: adaptive walks on the
encoded landscape perform best.

\subsection{Conclusion}

We have examined the role of encodings in arriving at optimal solutions to
NP-complete problems: we have constructed encodings for three examples,
\emph{viz.} the NPP, Spin-Glass and TSP problems, and demonstrated that the
choice of a good encoding can indeed help optimization. In the examples we
have chosen, the benefits arise primarily as a result of the enrichment of
low-energy solutions. A secondary effect in some but not all encodings
considered here is the introduction of a high degree of neutrality. The
latter enables a diffusion-like mode of search that can be much more
efficient than the combination of fast hill-climbing and exponentially rare
jumps from local optima. The two criteria, (1) selective enrichment of low
energy states and, where possible, (2) increase of local degeneracy, can
guide the construction of alternative encodings explicitly making use of
\emph{a priori} knowledge on the mathematical structure of optimization
problem. The qualitative understanding of the effect of encodings on
landscape structures in particular resolves apparently conflicting ``design
guidelines'' for the construction of evolutionary algorithms.

  The beneficial effects of enriching encodings immediately pose the
  question whether there is a generic way in which they can be constructed.
  The constructions for the NPP and TSP encodings suggest one rather
  general design principle. Suppose there is a natural way of decomposing a
  solution $x$ of the original problem into partial solutions. We can think
  of a partial solution $\xi$ as the set of all solutions that have a
  particular property. In the TSP example, $\xi$ refers to a set of
  solutions in which a certain list $A$ of cities appears as an
  uninterrupted interval. Now we choose the encoding $y$ so that it has an
  \emph{interpretation} as a collection $\Xi(y)$ of partial solutions. A
  deterministic optimization heuristic can now be used to determine a good
  solution $x^*(\Xi(y))$. In the case of the TSP, $\Xi(y)$ corresponds to a
  set of constrained tours from which we choose by a greedy solution.
  Alternatively, $\Xi(y)$ may over-specify a solution, in which case the
  optimization procedure would attempt to extract an optimal subset of
  $\Xi'\subseteq \Xi(y)$ so that $\bigcap_{\xi\in\Xi'}\xi$ contains a valid
  solution $x^*$. In either case, $\alpha:y\mapsto x^*$ is an encoding that
  is likely to favour low-energy states. It is not obvious, however, that the
  spanning-tree encoding for max-cut can also be understood as a
  combination of partial solutions. It remains an important question for
  future research to derive necessary and sufficient conditions under which
  optimized combinations of partial solutions indeed guarantee that the
  encoding is enriching.

\section*{Methods}
\subsection*{Landscapes and encoding}

A finite discrete energy landscape $(X,L,f)$ consists of a finite set of
configurations $X$ endowed with an adjacency structure $L$ and with a
function $f:X \rightarrow \mathbb{R}$ called energy, and hence $-f$
fitness. The global minima of $f$ are called ground states.
$L$ is a set of unordered tuples in $X$, thus $(X,L)$ is a simple
undirected graph. Let $(Y,M)$ be another simple graph and consider a
mapping $\alpha:Y \rightarrow X\cup\{\varnothing\}$, which we call an
encoding of $X$.  Then $(Y,M,f \circ \alpha)$ is again a
landscape. (If we include states in $Y$ that do not encode feasible
  solutions we assign them infinite energy, i.e., $f\circ\alpha(y)=+\infty$
  if $\alpha(y)=\varnothing$.)  The encoding is \emph{complete} if
$\alpha$ is surjective, i.e., if every $x\in X$ is encoded by at least one
vertex of $y\in Y$. Both landscapes then describe the same optimization
problem. In the language of evolutionary computation, $(Y,M)$ is the
genotype space, while $(X,L)$ is the phenotype space corresponding to the
``direct encoding'' of the problem. With this notation fixed, our problem
reduces to understanding the differences between the genotypic landscape
$(Y,M,f \circ \alpha)$ and the phenotypic landscape $(X,L,f)$ w.r.t.\
optimization dynamics.

\subsection*{Test Instances} 

Random instances fox max-cut (spin glass) are generated as standard random
graphs \cite{Bollobas:01} with parameter $p=0.5$: each potential edge is
present or absent with equal probability, independent from other
edges. Distances $d_{ij}=d_{ji}$ for the symmetric TSP and numbers $a_i$
for NPP are drawn independently from the uniform distribution on the
interval $[0,1]$.

\subsection*{Enrichment factor and Density of States}

The \emph{enrichment factor} $r(h)$ can be obtained directly from the
  cumulative densities of states of the two landscapes:
  \begin{equation}
    r(h)= Q_{f \circ \alpha} ( Q_f ^{-1} (h))~.
  \end{equation}
  This expression is a well-defined function for arguments $h \in [0,1]$
  because $Q_{f \circ \alpha}$ only changes value where $Q_f$ also does.
  For ground state energy $\eta_0$, the enrichment of the ground state is
  $Q_{f \circ \alpha}(\eta_0) / Q_f(\eta_0)$.

The results in Figure~\ref{fig:enrichment}(a-c) are obtained by sampling 
$2\times 10^7$ uniformly drawn states each from the original states $X$ and the
prepartitionings $Y$ for the traveling salesman. For the two other problems, the
density of states of the original landscapes is exact by complete enumeration.
For the spin glass also, the density of states for $Y$ is exact from calculation
based on the matrix-tree theorem.  For number partitioning, $2^n$ samples in $Y$
are drawn at random.

The enrichment of the ground state, Figure~\ref{fig:enrichment}(d), is
an average over 100 realizations for each problem type and size $n$. For each
realization of number partitioning and max-cut, $2^n$ uniform samples in
$Y$ are taken; the ground state energy itself is obtained by complete
enumeration of $X$. For each realization of the traveling salesman problem,
$10^9$ uniform samples are taken in $Y$; the ground state energy is computed
with the Karp-Held algorithm \cite{Held:62}.

\section*{Acknowledgments}
K.K.\ and P.F.S.\ thank Volkswagenstiftung for financial support.
\bibliography{encoding_o}

\end{document}